\documentclass{cmfnr}
\usepackage[cp1251]{inputenc}
\usepackage[russian]{babel}
\usepackage{graphicx}
\usepackage{epstopdf}

\title[Меры Гиббса для НС-модели на дереве Кэли]{Периодические меры Гиббса для НС-модели с двумя состояниями на дереве Кэли} 
\subjclass{517.98} 
\author{У.А.Розиков, Р.М.Хакимов, М.Т.Махаммадалиев } 
\address{Институт математика имени В.И.Романовского, Ташкент, Узбекистан.} 
\email{rozikovu@yandex.ru}
\address{Наманганский государственный университет, Наманган, Узбекистан.} 
\email{rustam-7102@rambler.ru}
\address{Наманганский государственный университет, Наманган, Узбекистан.} 
\email{mmtmuxtor93@mail.ru}

\theoremstyle{plain} 
\newtheorem{theorem}{Теорема}

\newtheorem{lemma}{Лемма}

\theoremstyle{definition} 
\newtheorem{definition}{Определение}

\numberwithin{equation}{section}
\numberwithin{theorem}{section}
\numberwithin{lemma}{section}
\numberwithin{definition}{section}
\numberwithin{remark}{section}
\numberwithin{proposition}{section}
\numberwithin{corollary}{section}
\begin{document}
\begin{abstract}
В данной статье изучается Hard-Core (НС) модель с двумя состояниями и активностью $\lambda>0$ на дереве Кэли порядка $k\geq 2$.
Известно, что существуют $\lambda_{\rm cr}$, $\lambda^0_{\rm cr}$ , $\lambda'_{\rm cr}$ такие, что

-при $\lambda\leq \lambda_{\rm cr}$ для этой модели существует единственная мера Гиббса $\mu^*$, которая является
трансляционно-инвариантной. Мера $\mu^*$ является крайней при $\lambda<\lambda^0_{\rm cr}$  и не крайней при $\lambda>\lambda'_{\rm cr}$.

-при $\lambda>\lambda_{\rm cr}$ существуют равно три два-периодические меры Гиббса, одна из которых является $\mu^*$, две остальные
являются не трансляционно-инвариантными и всегда крайними.

Крайность этих периодических мер была доказана с помощью максимальности и минимальности
соответствующих решений некоторого уравнения, обеспечивающего
согласованность этих мер. В данной работе мы дадим краткий обзор известных мер Гиббса для НС-модели и
альтернативное доказательство крайности два-периодических мер при $k=2,3$. Наше доказательство основано на метод реконструкции
на дереве.
\end{abstract}
\maketitle
\tableofcontents


\section{Введение}
\label{sec1}

Решения проблем, возникающих в результате исследований при изучении
термодинамических свойств физических и биологических систем, в основном
приводятся к задачам теории мер Гиббса. Мера Гиббса-это фундаментальный закон, определяющий
вероятность микроскопического состояния данной физической системы. Известно, что каждой
мере Гиббса сопоставляется одна фаза физической системы и если мера Гиббса не единственна,
то говорят, что существует фазавый переход. Для достаточно широкого класса гамильтонианов
известно, что множество всех предельных мер Гиббса (соответствующих данному гамильтониану)
образует непустое выпуклое компактное подмножество в множестве всех вероятностных мер (см. например \cite{6}, \cite {Si}, \cite{Pr})
и каждая точка этого выпуклого множества однозначно разлагается по его крайним точкам. В связи
с этим особый интерес представляет описание всех крайних точек этого выпуклого множества, т. е. крайних мер Гиббса.

Определение меры Гиббса и других понятий, связанных с теорией мер Гиббса, можно
найти, например, в работах \cite{6}, \cite{Si}, \cite{Pr}, \cite{Rb}. Несмотря на многочисленные работы, посвященных изучению мер Гиббса, ни для одной модели не было получено полное описание всех предельных мер Гиббса.
Относительно других моделей для модели Изинга на дереве Кэли эта задача изучена достаточно полностью. Так,
например, в работе \cite{Bl} построено несчетное множество крайних
гиббсовских мер, а в работе \cite{Bl1} найдено необходимое и
достаточное условие крайности неупорядоченной фазы модели Изинга
на дереве Кэли.

В работе \cite{Maz} Мазелью и Суховым была введена и изучена НС-модель (жесткий диск, жесткая сердцевина) на $d$-мерной решетке $ \mathbb Z ^ d$.
Работы \cite{RKh2}, \cite{RKh}, \cite{Kh}-\cite{KhRM}, \cite{KhM}, \cite{Mar}-\cite{7} посвящены изучению (слабо) периодических мер Гиббса для HC-модели с двумя состояниями на дереве Кэли. В работе \cite{7} была доказана единственность трансляционно-инвариантной меры и не единственность
периодических мер Гиббса для НС-модели. Также в \cite{7}
(соответственно в \cite{Mar}) найдено достаточное условие на
параметры НС-модели, при котором трансляционно-инвариантная мера Гиббса является не крайней
(соответственно крайней). В работе \cite{RKh2} расширена область крайности этой меры. Работа \cite{RKh} посвящена изучению слабо периодических мер Гиббса для HC-модели для нормального делителя индекса два и при некоторых условиях на параметры показана единственность слабо периодической меры Гиббса, а в работе \cite{Kh} дано полное описание слабо периодических мер Гиббса для HC-модели при любых значениях параметров в случае нормального делителя индекса два.  Для ознакомления с другими свойствами НС-модели (и их обобщения)
на дереве Кэли см. Главу 7 монографии \cite{Rb}.

В данной работе изучается НС-модель с двумя состояниями на дереве Кэли. Даны новые доказательства утверждения, что на дереве Кэли порядка два и три существуют ровно три $G_k^{(2)}$-периодические меры Гиббса. При этом найдены явные виды $G_k^{(2)}$-периодических (не трансляционно-инвариантных) мер Гиббса на дереве Кэли порядка два и три. Доказано, что эти меры являются крайними.

\section{Предварительные сведения и известные факты}\

{\bf 1. Дерево Кэли.} Пусть $\Gamma^k=(V,L,i)$ есть дерево Кэли порядка $ k\geq 1$, где $V$ есть множество
вершин $\Gamma^k$, $L-$ множество его ребер и $i-$ функция
инцидентности, сопоставляющая каждому ребру $l\in L$ его концевые
точки $x, y \in V$. Если $i (l) = \{ x, y \} $, то $x$ и $y$
называются  {\it ближайшими соседями вершины} и обозначается $l =
\langle x,y\rangle $. Пусть $d(x,y), x, y \in V$ есть расстояние между вершинами
$x, y$, т.е. количество ребер кратчайшей пути, соединяющей $x$ и $y$.

Для фиксированного $x^0\in V$ обозначим
$$ W_n =\{x\in V | \ d (x, x^0) =n \}, \  V_n = \cup_{j=0}^{n}W_j.$$
Для $x\in W_{n}$ обозначим (множество прямых потомков вершины $x$)
$$ S(x)=\{y\in{W_{n+1}}:d(x,y)=1\}.$$

{\bf 2. Допустимые конфигурации.} Пусть $\Phi=\{0,1\}$ и $\sigma\in\Phi^V-$ конфигурация, то есть
$\sigma=\{\sigma(x)\in \Phi: x\in V\}$, где $\sigma (x)=1$
означает, что вершина $x$ на дереве Кэли занята, а $\sigma (x)=0$
означает, что она свободна. Конфигурация $\sigma$ называется
допустимой, если $\sigma (x)\sigma (y)=0$ для любых соседних
$\langle x,y \rangle $ из $V$ ($V_n $ или $W_n$, соответственно) и
обозначим множество таких конфигураций через $\Omega$
($\Omega_{V_n}$ и $\Omega_{W_n}).$ Ясно, что
$\Omega\subset\Phi^V.$

Объединение конфигураций $\sigma_{n-1}\in\Phi ^ {V_{n-1}}$ и $\omega_n\in\Phi
^ {W_{n}}$ определяется следующей формулой
$$
\sigma_{n-1}\vee\omega_n=\{\{\sigma_{n-1}(x), x\in V_{n-1}\},
\{\omega_n(y), y\in W_n\}\}.
$$

{\bf 3. Мера Гиббса.}  Гамильтониан HC-модели определяется по формуле
$$H(\sigma)= J \sum\limits_{x\in{V}}{\sigma(x),} \ \sigma \in\Omega,$$
где $J\in R$.

Пусть $\mathbf{B}$ есть $\sigma$-алгебра, порожденная
цилиндрическими подмножествами $\Omega.$ Для любого $n$ обозначим
через $\mathbf{B}_{V_n}=\{\sigma\in\Omega:
\sigma|_{V_n}=\sigma_n\}$ подалгебру $\mathbf{B},$ где
$\sigma|_{V_n}-$ сужение $\sigma$ на $V_n,$ $\sigma_n: x\in V_n
\mapsto \sigma_n(x)-$ допустимая конфигурация в $V_n.$

\begin{definition}\label{q1}
Для $\lambda >0$ НС-мера Гиббса есть
вероятностная мера $\mu$ на $(\Omega , \textbf{B})$ такая, что для
любого $n$ и $\sigma_n\in \Omega_{V_n}$
$$
\mu \{\sigma \in \Omega:\sigma|_{V_n}=\sigma_n\}=
\int_{\Omega}\mu(d\omega)P_n(\sigma_n|\omega_{W_{n+1}}),
$$
где
$$
P_n(\sigma_n|\omega_{W_{n+1}})=\frac{e^{-H(\sigma_n)}}{Z_{n}
(\lambda ; \omega |_{W_{n+1}})}\textbf{1}(\sigma_n \vee \omega
|_{W_{n+1}}\in\Omega_{V_{n+1}}).
$$

Здесь $Z_n(\lambda ; \omega|_{W_{n+1}})-$ нормировочный множитель с
граничным условием $\omega|_{W_n}$:
$$
Z_n (\lambda ; \omega|_{W_{n+1}})=\sum_{\widetilde{\sigma}_n \in
\Omega_{V_n}}
e^{-H(\widetilde{\sigma}_n)}\textbf{1}(\widetilde{\sigma}_n\vee
\omega|_{W_{n+1}}\in \Omega_{V_{n+1}}).
$$
\end{definition}

Для $\sigma_n\in\Omega_{V_n}$ положим
$$\#\sigma_n=\sum\limits_{x\in V_n}{\mathbf 1}(\sigma_n(x)\geq 1)$$
число занятых вершин в $\sigma_n$.

Пусть $z:x\mapsto z_x=(z_{0,x}, z_{1,x}) \in R^2_+$
векторнозначная функция на $V$. Для $n=1,2,\ldots$ и $\lambda>0$
рассмотрим вероятностную меру $\mu^{(n)}$ на $\Omega_{V_n}$,
определяемую как
\begin{equation}\label{e1}
\mu^{(n)}(\sigma_n)=\frac{1}{Z_n}\lambda^{\#\sigma_n} \prod_{x\in
W_n}z_{\sigma(x),x}.
\end{equation}
Здесь $Z_n-$ нормирующий делитель:
$$Z_n=\sum_{{\widetilde\sigma}_n\in\Omega_{V_n}}
\lambda^{\#{\widetilde\sigma}_n}\prod_{x\in W_n}
z_{{\widetilde\sigma}(x),x}.$$

Говорят, что последовательность вероятностных мер $\mu^{(n)}$ является
согласованной, если для любых $n\geq 1$ и
$\sigma_{n-1}\in\Omega_{V_{n-1}}$:
\begin{equation}\label{e2}
\sum_{\omega_n\in\Omega_{W_n}}
\mu^{(n)}(\sigma_{n-1}\vee\omega_n){\mathbf 1}(
\sigma_{n-1}\vee\omega_n\in\Omega_{V_n})=
\mu^{(n-1)}(\sigma_{n-1}).
\end{equation}
В этом случае существует единственная мера $\mu$ на $(\Omega,
\textbf{B})$ такая, что для всех $n$ и $\sigma_n\in \Omega_{V_n}$
$$\mu(\{\sigma|_{V_n}=\sigma_n\})=\mu^{(n)}(\sigma_n).$$

\begin{definition}\label{q2}
Мера $\mu$, являющейся пределом последовательности $\mu^{(n)}$, определенной формулой
(\ref{e1}) с условием согласованности (\ref{e2}), называется HC-\textit{мерой Гиббса} с $\lambda>0$,
\textit{соответ-ствующей функции} $z:\,x\in V \setminus\{x^0\}\mapsto z_x$. \
\end{definition}

Известно, что существует взаимнооднозначное соответствие между
множеством $V$ вершин дерева Кэли порядка $k\geq 1$ и группой
$G_k$, являющейся свободным произведением $k+1$ циклических групп
второго порядка с образующими $a_1,...,a_{k+1}$, соответственно
(см. \cite{1}). Поэтому множество $V$ можно отождествлять
c множеством $G_k$.

Известно \cite{7}, что каждой HC- мере Гиббса для
HC-модели на дереве Кэли можно сопоставлять совокупность величин $z=\{z_x, x\in
G_k \},$ удовлетворяющих
\begin{equation}\label{e3}
z_x=\prod_{y \in S(x)}(1+\lambda z_y)^{-1},
\end{equation}
где $\lambda=e^{J_1}>0-$ параметр, $J_1=-J\beta$, $\beta={1\over T}$, $T>0-$температура.\

Пусть $\widehat{G}_k-$ подгруппа группы $G_k$.

\begin{definition}\label{q3} Совокупность величин $z=\{z_x,x\in G_k\}$
называется $ \widehat{G}_k$-периодической, если  $z_{yx}=z_x$ для
$\forall x\in G_k, y\in\widehat{G}_k.$\

$G_k$-периодические совокупности называются
трансляционно-инвариантными.
\end{definition}

Для любого $x\in G_k $ множество $\{y\in G_k: \langle
x,y\rangle\}\setminus S(x)$ имеет единственный элемент, которого
обозначим через $x_{\downarrow}$ (см.\cite{RR}).

Пусть $G_k/\widehat{G}_k=\{H_1,...,H_r\}$ фактор группа, где
$\widehat{G}_k$ есть нормальный делитель индекса $r\geq 1.$

\begin{definition}\label{q4} Совокупность величин $z=\{z_x, x\in G_k\}$
называется $\widehat{G}_k$-слабо периодической, если
$z_x=z_{ij}$ при $x\in H_i, x_{\downarrow}\in H_j$ для $\forall
x\in G_k$.\
\end{definition}

\begin{definition}\label{q5} Мера $\mu$ называется
$\widehat{G}_k$-(слабо) периодической, если она соответствует
$\widehat{G}_k$-(слабо) периодической совокупности величин $z$.\
\end{definition}
Заметим, что мера $\mu$ является трансляционно-инвариантной, если она соответствует
$G_k$-периодической совокупности величин $z$.\\

{\bf 4. Известные теоремы.}

Известно, что для каждого $\beta > 0 $ меры Гиббса образуют
непустое выпуклое компактное множество $\mathbf {G} $ в пространстве всех
вероятностных мер на $\Omega $, обеспеченным слабой топологией \cite[Chapter~7]{6}.

Обозначим через ${\rm ex}\mathbf {G}$  множество всех крайних мер (точек) в $\mathbf {G}$.
Заметим, что (см. \cite[Теорема ~ (12.6)]{6}) каждой крайней мере
$\mu\in ex\mathbf {G} $ соответствует некоторое решение функционального уравнения (\ref{e3}).
Но обратное неверно: могут существовать решения, не определяющие крайние меры.
Теперь приведем несколько решений уравнения (\ref{e3}) и
условия крайности соответствующих мер Гиббса.

\begin{theorem}\label{te1} \cite{RKh} Для любого нормального делителя
${\mathcal G}\subset G_k$ всякая $\mathcal G$-периодическая мера
Гиббса НС-модели является либо трансляционно-инвариантной,
либо  $G^{(2)}_k$-периодической мерой Гиббса, где
$$G^{(2)}_k=\{x\in G_k: |x|-\emph{четное число}\}.$$\
\end{theorem}
\begin{theorem}\label{te2} \cite{7} \begin{itemize}
\item Для НС модели при $k\geq2$ и $\lambda>0$
трансляционно-инвариантная мера  Гиббса $\mu^*$ единственна.
\item Для $k\geq 2$ и
\begin{equation}\label{sr5.1}
\lambda>{1\over \sqrt{k}-1}\left({\sqrt{k}\over \sqrt{k}-1}\right)^k
\end{equation}
мера $\mu^*$ не является крайней.
\item Для любого $\epsilon > 0$ существует $k_0$ такое, что  мера $\mu^*$
не является крайней для всех $k\geq k_0$ и
\begin{equation}\label{sr5.2}
\lambda>e^{1+\epsilon}\ln k(\ln k+ \ln\ln k+1+\epsilon).
\end{equation}
\item Если для некоторых $k$ и $\lambda^0$ мера $\mu^*$
не является крайней, то она остается не крайней для этого
$k$ и для всех $\lambda > \lambda^0$.
\end{itemize}
\end{theorem}

Следующая теорема доказана в \cite{Mar}.

\begin{theorem}\label{srt5.4} Для $\lambda = 1$ мера $\mu^*$ является крайней для всех $k$.
\end{theorem}

При доказательстве крайности некоторой меры Гиббса $\mu$ обычно принимается алгоритм
реконструкции, предложенного и изученного в \cite{Mo} (см. также главу 4 \cite{Rb}).
В этом алгоритме рекурсивно присваивается значение 1 вершине $x$, если все $y\in S (x)$ имеют значения 0, и значение 0
в противном случае. Говорят, что реконструкция возможна, если при данной конфигурации $\sigma_n\in\Omega_{W_ n}$
возможно определить значения 0 или 1 для начальной точки $x^0 $.
Известно, что возможность реконструкции эквивалентна не крайности меры $\mu$.
Точнее, рассматривается 0, 1-значная цепь Маркова, порожденная мерой
$\mu$ на пути дерева. Пусть матрица вероятностных переходов для цепи есть
${\bf P}=(p_{ij})_{i,j=0,1}$.

В \cite{Mos2} доказано, что реконструкция невозможна, если
$${(p_{00}-p_{10})^2\over \min\{p_{00}+p_{10},\ \ p_{01}+p_{11}\}}\leq {1\over k}.$$

В работе \cite{Mar} эта оценка улучшена следующим образом.

\begin{theorem}\label{martin}
Реконструкция невозможна, если
\begin{equation}\label{mar1}
\left(\sqrt{p_{00}p_{11}}-\sqrt{p_{01}p_{10}}\right)^2\leq{1\over k}.
\end{equation}
\end{theorem}

Основным результатом \cite{RKh2} является следующая теорема.

\begin{theorem}\label{RK} При $k\geq2$ и $\lambda\in (0,\lambda_*)$
мера $\mu^{*}$ является крайней, где
\begin{equation}\label{l*}
\lambda_*=\lambda_*(k)={1\over t_*^k}\left({1\over t_*}-1\right),
\end{equation}
и $t_*\in (0,1)$ является единственным решением уравнения
\begin{equation}\label{z0}
t^{k+1}-k t^2+(2k-1)t-k+1=0.
\end{equation}
\end{theorem}

Теперь дадим полное описание $G^{(2)}_k$-периодических мер Гиббса на
дереве Кэли порядка два и три. Они соответствуют совокупности величин
 \begin{equation}\label{pp}z_x=\left\{%
\begin{array}{ll}
   z_1, \ $ если $ x \in G^{(2)}_k $,$ \\
   z_2, \ $ если $ x \in G_k\setminus G^{(2)}_k. $ $ \\
\end{array}%
\right. \end{equation}

Отсюда, в силу (\ref{e3}) имеем
\begin{equation}\label{e4}
\left\{%
\begin{array}{ll}
    z_{1}=\frac{1}{(1+\lambda z_2)^{k}} \\[3 mm]
    z_{2}=\frac{1}{(1+\lambda z_1)^{k}}. \\
    \end{array}%
\right.
\end{equation}

Известна следующая теорема.

\begin{theorem}\label{te3}\cite{7} \textit{Для НС модели при}
\begin{equation}\label{e5}
\lambda\leq\lambda_{cr}=(k-1)^{-1}\left(\frac{k}{k-1}\right)^k.
\end{equation}
\textit{существует ровно одна $G^{(2)}_k$-периодическая мера Гиббса $\mu_0$, которая совпадает
с единственной трансляционно-инвариантной мерой Гиббса $\mu^*$ и при $\lambda>\lambda_{cr}$
существуют ровно три $G^{(2)}_k$-периодические меры Гиббса $\mu_0, \mu_1, \mu_2$, где мера $\mu_0$ является
трансляционно-инвариантной, а меры $\mu_1$ и $\mu_2$ являются $G^{(2)}_k$-периодическими (не трансляционно-инвариантными).}
\end{theorem}

\textbf{Слабо периодические меры.} В работе \cite{Kh} были изучены слабо периодические меры Гиббса для любых нормальных делителей индекса два и доказана, что такая мера единственна. Более того, она совпадает с единственной трансляционно-инвариантной мерой Гиббса.

Пусть $A\subset\{1,2,...,k+1\}$ и $H_A=\{x\in
G_k:\sum\limits_{i\in A}w_x(a_i)-$ четное число$ \}$, где
$w_x(a_i)-$ число буквы $a_i$ в слове $x\in G_k$
 и
$G_k^{(4)}=H_A\cap G_k^{(2)}-$ нормальный делитель индекса 4.

Тогда, в силу (\ref{e3}), $G_k^{(4)}$-слабо периодические меры соответствуют решениям следущей системы уравнения

\begin{equation}\label{e34}
\left\{%
\begin{array}{ll}
    z_{1}=\frac{(1+\lambda z_{3})^k}{((1+\lambda z_{3})^{k/i}+\lambda z_{4}^{1-1/i})^i}\cdot\frac{1}{(1+\lambda z_2)^{k-i}} \\
    \\
    z_{2}=\frac{(1+\lambda z_{4})^k}{((1+\lambda z_{4})^{k/i}+\lambda z_{3}^{1-1/i})^i}\cdot\frac{1}{(1+\lambda z_1)^{k-i}} \\
    \\
    z_{3}=\frac{(1+\lambda z_1)^k}{((1+\lambda z_1)^{k/i}+\lambda z_2^{1-1/i})^i}\cdot\frac{1}{(1+\lambda z_{4})^{k-i}} \\
    \\
    z_{4}=\frac{(1+\lambda z_2)^k}{((1+\lambda z_2)^{k/i}+\lambda z_1^{1-1/i})^i}\cdot\frac{1}{(1+\lambda z_{3})^{k-i}} \\
\end{array}%
\right.
\end{equation}
Здесь $i=|A|-$ мощность множества $A$.

Рассмотрим отображение $W:R^4 \rightarrow R^4,$ определенное
следующим образом:

$$
\left\{%
\begin{array}{ll}
    z_{1}^{'}=\frac{(1+\lambda z_{3})^k}{((1+\lambda z_{3})^{k/i}+\lambda z_{4}^{1-1/i})^i}\cdot\frac{1}{(1+\lambda z_2)^{k-i}}
    \\[3mm]
    z_{2}^{'}=\frac{(1+\lambda z_{4})^k}{((1+\lambda z_{4})^{k/i}+\lambda z_{3}^{1-1/i})^i}\cdot\frac{1}{(1+\lambda z_1)^{k-i}}
    \\[3mm]
    z_{3}^{'}=\frac{(1+\lambda z_1)^k}{((1+\lambda z_1)^{k/i}+\lambda z_2^{1-1/i})^i}\cdot\frac{1}{(1+\lambda z_{4})^{k-i}}
    \\[3mm]
    z_{4}^{'}=\frac{(1+\lambda z_2)^k}{((1+\lambda z_2)^{k/i}+\lambda z_1^{1-1/i})^i}\cdot\frac{1}{(1+\lambda z_{3})^{k-i}} \\
\end{array}%
\right.
$$

Заметим, что (\ref{e34}) есть уравнение $z=W(z).$ Чтобы решить систему уравнений (\ref{e34}), надо
найти неподвижные точки отображения $z^{'}=W(z).$

Известны следующие леммы.

\begin{lemma}\label{ru11} \cite{Kh1} \textit{ Отображение $W$ имеет инвариантные
множества следующих видов:}
$$I_1=\{(z_1, z_2, z_{3}, z_{4}) \in R^4: z_1=z_2=z_{3}=z_{4}\}, \ \ I_2=\{(z_1, z_2, z_{3}, z_{4})\in R^4: z_1=z_{3}, \ z_2=z_{4}\},$$
$$I_3=\{(z_1, z_2, z_{3}, z_{4}) \in R^4: z_1=z_2, z_{3}=z_{4}\}, \ \ I_4=\{(z_1, z_2, z_{3}, z_{4})\in R^4: z_1=z_{4}, \ z_2=z_{3}\}.$$
\end{lemma}
\begin{lemma}\label{ru12} \cite{Kh1} \textit{Если на инвариантных множествах
$I_2, I_3, I_4$ существуют слабо периодические меры Гиббса, то они
являются либо трансляционно-инвариантными, либо слабо
периодическими (не периодическими).}
\end{lemma}\

В работах \cite{Kh1}, \cite{Kh2}, \cite{KhRM} и \cite{KhM} были изучены слабо периодические меры Гиббса для HC-модели в случае нормального
делителя индекса четыре на некоторых инвариантах. В частности, были доказаны утверждения следующей теоремы.

\begin{theorem}\label{te22} Для HC-модели в случае нормального
делителя индекса четыре верны следующие утверждения:\begin{itemize}

\item Пусть $k=2$, $\lambda_{cr}=4$ и $i=1$ или $i=2$. Тогда на $I_2$ при
$\lambda<\lambda_{cr}$ существует одна слабо периодическая мера
Гиббса, которая является трансляционно-инвариантной, при
$\lambda=\lambda_{cr}$ существуют две слабо периодические меры
Гиббса, одна из которых является трансляционно-инвариантной,
другая слабо периодической (не периодической) и при
$\lambda>\lambda_{cr}$ существуют ровно две слабо
периодические (не периодические) меры Гиббса.
\item Пусть $k=3, i=1, \lambda_{cr}={27\over16}$. Тогда для HC-модели в случае нормального
делителя индекса четыре при $\lambda\leq\lambda_{cr}$ существует одна слабо
периодическая мера Гиббса (соответствующая совокупности величин из $I_2$), которая является
трансляционно-инвариантной, а при $\lambda>\lambda_{cr}$ существуют ровно три слабо периодические меры Гиббса (соответствующие совокупности величин из $I_2$), одна из которых является трансляционно-инвариантной, а две другие слабо периодическими
(не периодическими)
\item При $k\geq6$, $i=1$ и $\lambda\in(\lambda^{-}(k), \lambda^{+}(k))$ существуют не менее трех слабо периодических мер Гиббса, соответствующих совокупности величин из $I_4$. При этом одна из них является трансляционно-инвариантной, другие слабо периодическими (не периодическими) мерами Гиббса, где
    $$s^{\pm}:=s^{\pm}(k)=\frac{k-3\pm\sqrt{k^2-6k+1}}{4},$$
$$\lambda^{\pm}:=\lambda^{\pm}(k)=(s^{\pm}+1)^ks^{\pm}.$$
\end{itemize}
\end{theorem}

\textbf{Замечание.} По Лемме \ref{ru12} ясно, что существующие слабо периодические меры Гиббса в Теореме \ref{te22} отличаются от периодических.

\section{Условия крайности периодических мер}\

Обозначим $f(x)=\frac{1}{(1+\lambda x)^{k}}$.

Следующая лемма очевидна.

\begin{lemma}\label{ru} Если  $(x_0, y_0)$ является решением системы уравнений
\begin{equation}\label{e6}
\left\{%
\begin{array}{ll}
    x=f(y) \\
    y=f(x), \\
    \end{array}%
\right.
\end{equation}
то $(y_0, x_0)$ тоже является решением системы уравнений (\ref{e6}).
\end{lemma}

В частности, из леммы следует, что если существует решение $(x_0, y_0)$ ($x_0\neq y_0$),
то (\ref{e6}) имеет более одного решения.

При изучении крайности нам необходимы явные виды решений,
соответствующих мерам $\mu_1$ и $\mu_2$. Теперь мы найдем явные виды решений
при $k=2, 3$.

\textbf{Случай $k=2$.} Перепишем систему уравнений (\ref{e4}) при $k=2$:
\begin{equation}\label{e7}
\left\{%
\begin{array}{ll}
    \sqrt{z_{1}}+\lambda z_2\sqrt{z_{1}}=1 \\[2 mm]
    \sqrt{z_{2}}+\lambda z_1\sqrt{z_{2}}=1. \\
    \end{array}%
\right.
\end{equation}
Введя обозначения $\sqrt{z_{1}}=x$ и $\sqrt{z_{2}}=y$, перепишем (\ref{e7})
следующим образом:
\begin{equation}\label{e8}
\left\{%
\begin{array}{ll}
    x+\lambda xy^2=1 \\[2 mm]
    y+\lambda yx^2=1. \\
    \end{array}%
\right.
\end{equation}
В этой системе уравнений, вычтев из первого уравнения$-$ второе, получим
$$(x-y)(1-\lambda xy)=0.$$
Отсюда $x=y$ или
$$\lambda xy=1.$$
В случае $x=y$ получим решение, соответствующее единственной
трансляционно-инвариантной мере Гиббса $\mu_0$ при $\lambda>0$. Это решение имеет явный вид.
Кроме того, в работе \cite{Kh2} была изучена крайность трансляционно-инвариантной
меры Гиббса, соответствующей этому решению.

Пусть $x\neq y$ и $\lambda xy=1.$ В этом случае, из (\ref{e8}) получим
квадратное уравнение
$$\lambda x^2-\lambda x+1=0,$$
решения которого имеют вид
$$x_1=\frac{\lambda+\sqrt{\lambda^2-4\lambda}}{2\lambda}, \ x_2=\frac{\lambda-\sqrt{\lambda^2-4\lambda}}{2\lambda}.$$
Ясно, что $\lambda>\lambda_{cr}=4$ и $x_1>0,\ x_2>0$.

Далее, из $\lambda xy=1$ получим
$$y_1=\frac{2}{\lambda+\sqrt{\lambda^2-4\lambda}}, \ y_2=\frac{2}{\lambda-\sqrt{\lambda^2-4\lambda}}.$$

Итак, для системы уравнений (\ref{e8}) имеем решения вида: $(x_1,y_1)$ и $(x_2,y_2)$. В силу Леммы \ref{ru}
$(y_1,x_1)$ и $(y_2,x_2)$ также являются решениями (\ref{e8}). Но не трудно заметить, что $x_1=y_2$ и $x_2=y_1$. Из всего сказанного следует, что
при $0<\lambda\leq\lambda_{cr}$ существует ровно одна $G^{(2)}_k$-периодическая мера Гиббса, которая совпадает
с единственной трансляционно-инвариантной мерой Гиббса $\mu_0$, а при $\lambda>\lambda_{cr}$
существуют ровно три $G^{(2)}_k$-периодические меры Гиббса $\mu_0, \mu_1, \mu_2$, где
меры $\mu_1, \mu_2$ соответствуют решениям $(x_1,y_1)$ и $(x_2,y_2)$,
соответственно и являются $G^{(2)}_k$-периодическими (не трансляционно-инвариантными).

\textbf{Случай $k=3$.} Перепишем систему уравнений (\ref{e4}) при $k=3$:
\begin{equation}\label{e8}
\left\{%
\begin{array}{ll}
    \sqrt[3]{z_{1}}+\lambda z_2\sqrt[3]{z_{1}}=1 \\[2 mm]
    \sqrt[3]{z_{2}}+\lambda z_1\sqrt[3]{z_{2}}=1. \\
    \end{array}%
\right.
\end{equation}
Введя обозначения $\sqrt[3]{z_{1}}=x$ и $\sqrt[3]{z_{2}}=y$, перепишем (\ref{e8})
следующим образом:
\begin{equation}\label{e9}
\left\{%
\begin{array}{ll}
    x+\lambda xy^3=1 \\[2 mm]
    y+\lambda yx^3=1. \\
    \end{array}%
\right.
\end{equation}
В этой системе уравнений, вычтев из первого уравнения$-$ второе, получим
$$(x-y)(1-\lambda xy(x+y))=0.$$
Отсюда $x=y$ или
$$\lambda xy(x+y)=1.$$
В случае $x=y$ получим решение, соответствующее единственной
трансляционно-инвариантной мере Гиббса $\mu^{(0)}$ при $\lambda>0$. В работе \cite{RKh2} была изучена крайность трансляционно-инвариантной
меры Гиббса, соответствующей этому решению.

Пусть $x\neq y$. Тогда $\lambda xy(x+y)=1.$ Используя последнее равенство, из первого уравнения (\ref{e9}) получим
уравнение
$$x^2+y^2+xy=x+y.$$
Введем обозначения $x+y=a$ и $xy=b$. Тогда $ab\lambda=1$ и $a^2-b=a$.
Отсюда имеем уравнение
$$a^3-a^2-\frac{1}{\lambda}=0,$$
решение которого по формуле Кардано имеет вид
$$a=\frac{1}{6}\sqrt[3]{\frac{12\sqrt{12\lambda+81}+8\lambda+108}{\lambda}}+\frac{2}{3}\sqrt[3]{\frac{\lambda}{12\sqrt{12\lambda+81}+8\lambda+108}}+\frac13.$$
С другой стороны, из $x+y=a$, $xy=b$ следует, что $x$ и $y$ являются решениями некоторого квадратного уравнения
$t^2-at+b=0$.
Используя равенство $b=\frac{1}{a\lambda}$, для решений этого квадратного уравнения будем иметь:
$$t_{1,2}=\frac{\lambda a^2\pm \sqrt{\lambda^2 a^4-4\lambda a}}{2\lambda a},$$
т.е.
$$t_1=\frac{\lambda a^2 - \sqrt{\lambda^2 a^4-4\lambda a}}{2\lambda a}=x,   \ t_2=\frac{\lambda a^2+ \sqrt{\lambda^2 a^4-4\lambda a}}{2\lambda a}=y.$$
С помощью компьютерного анализа можно увидеть, что дискриминант $D(\lambda)=a^4\lambda^2-4a\lambda>0$ при
$\lambda>\lambda_{cr}=\frac{27}{16}$ и $D(\lambda_{cr})=0$ (см. Рис.1). Заметим, что это критическое значение $\lambda_{cr}$
совпадает с значением данной в Теореме \ref{te3} при $k=3$.
\begin{center}
\includegraphics[width=6cm]{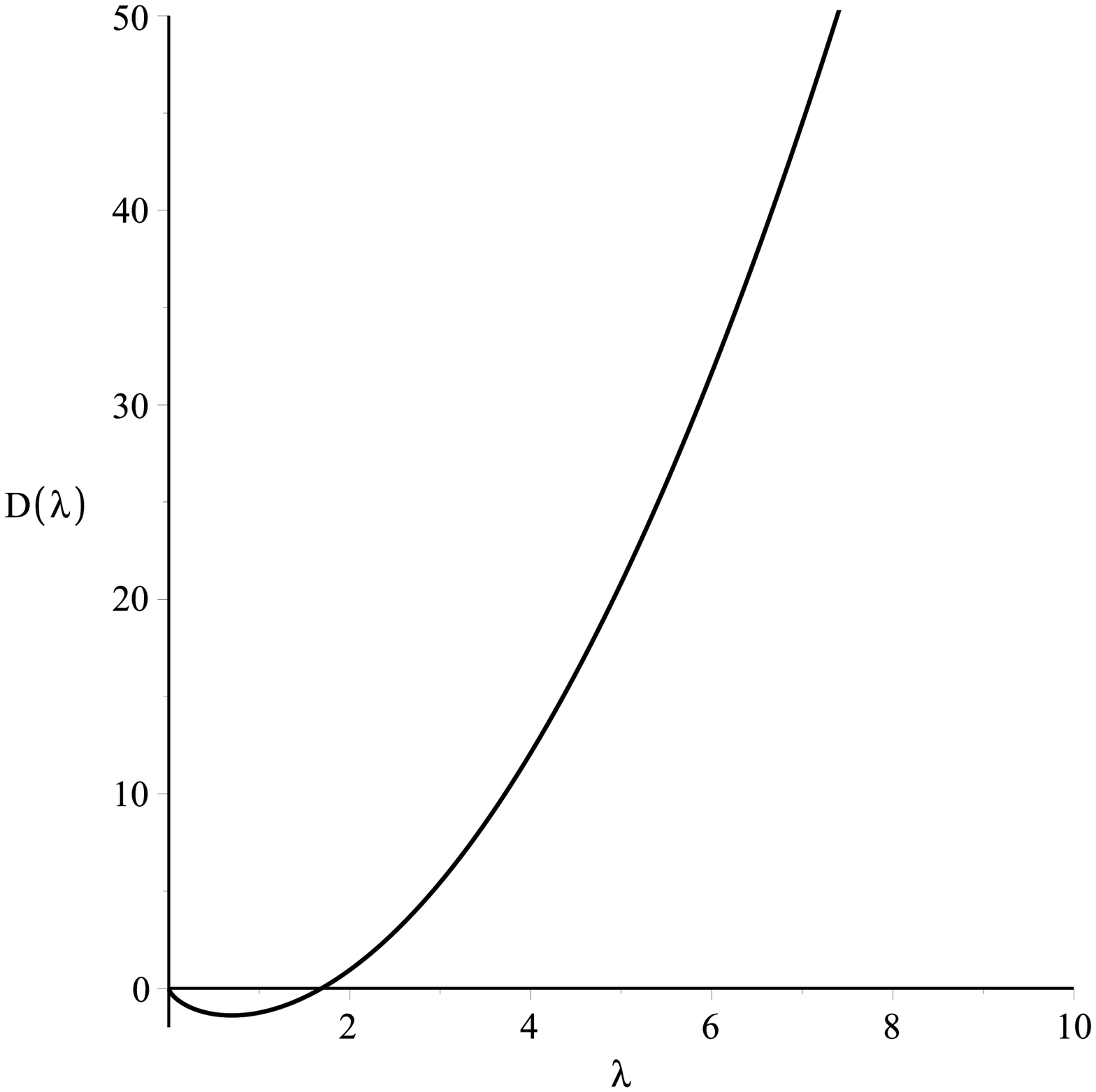}
\end{center}
\begin{center}{\footnotesize \noindent
 Рис. 1. График функции $D(\lambda)$.}\
\end{center}

Значит, система уравнений (\ref{e8}) имеет решение $(z_1,z_2)$, где
$$z_1=\left(\frac{\lambda a^2 - \sqrt{\lambda^2 a^4-4\lambda a}}{2\lambda a}\right)^3,
\ z_2=\left(\frac{\lambda a^2+ \sqrt{\lambda^2 a^4-4\lambda a}}{2\lambda a}\right)^3.$$
Но в силу симметрии $(z_2,z_1)$ тоже является решением (\ref{e8}). Итак, система уравнений (\ref{e8}) при $0<\lambda\leq\lambda_{cr}$ имеет единственное решение $(z,z)$, соответствующее единственной трансляционно-инвариантной мере $\mu^{(0)}$, а при $\lambda>\lambda_{cr}$ имеет три решения $(z,z)$, $(z_1,z_2)$ и $(z_2,z_1)$, которые соответствуют мерам $\mu^{(0)}, \mu^{(1)}, \mu^{(2)}$, где меры $\mu^{(1)}, \mu^{(2)}$ являются $G^{(2)}_k$-периодическими (не трансляционно-инвариантными).\\

\textbf{Крайность меры.}  Мы имеем $G^{(2)}_k$-периодические меры $\mu_1$ и $\mu_2$.
Чтобы изучить их (не) крайность воспользуемся методами из работ \cite{MSW}, \cite{KS}, \cite{Kr} и \cite{Kr1}
для трансляционно-инвариантных мер Гиббса.
Для каждой трансляционно-инвариантной меры рассматривается цепь
Маркова с состояниями $\{0,1\}$, индексированная на дереве Кэли, т.е.
предположим, что нам даны дерево Кэли с множеством вершин $V$,
вероятностная мера $\nu$, и матрица вероятностных переходов
$\mathbb P = \left(P_{ij}\right)$ на множестве $\{0,1\}$. Мы можем
построить дерево - индексированное цепью Маркова $X: V \to \{0,1\}$ путем выбора $X(x^0)$ в соответствии с $\nu$
и выбором $X(v)$, для каждой вершины $v\ne x^0$, используя вероятности перехода с
учетом значения его родителя, независимо от всего остального.
Так как трансляционно-инвариантные меры получаются при $z_1=z_2$, в (\ref{pp})
матрица $\mathbb P$ зависит только от $z_1$, более точно,
$$\mathbb P\equiv \mathbb P_{z_1}=\begin{pmatrix}\frac{1}{1+\lambda z_1} & \frac{\lambda
z_1}{1+\lambda z_1} \\1 & 0 \end{pmatrix}.$$
Но в случае периодических мер матрица $\mathbb P$ зависит от  $z_1$ и $z_2$, где $z_1\ne z_2$ и $(z_1, z_2)-$ решение системы уравнений (\ref{e4}).
Точнее, $\mathbb P\equiv \mathbb P_{z_1, z_2}=\mathbf{P_{\mu_1}}$ (соот. $\mathbf{P_{\mu_2}}$) вероятностных переходов $P_{il}$, определенную данной
периодической мерой Гиббса $\mu_1$ (соот. $\mu_2$). Заметим, что $\mathbf{P_{\mu_1}}$ есть произведение двух матрицы
вероятностных переходов:

\begin{equation}\label{rus2.7}\mathbf{P_{\mu_1}}=\mathbb P_{z_1}\mathbb P_{z_2}=\begin{pmatrix}\frac{1}{1+\lambda z_1} & \frac{\lambda
z_1}{1+\lambda z_1} \\1 & 0\end{pmatrix}\cdot
\begin{pmatrix}\frac{1}{1+\lambda z_2} & \frac{\lambda
z_2}{1+\lambda z_2} \\1 & 0\end{pmatrix}=
\begin{pmatrix}\frac{1+\lambda^2 z_1z_2+\lambda z_1}{(1+\lambda z_1)(1+\lambda z_2)} &
\frac{\lambda z_2}{(1+\lambda z_1)(1+\lambda z_2)}
\\\frac{1}{1+\lambda z_2} & \frac{\lambda z_2}{1+\lambda z_2}.\end{pmatrix}\end{equation}

Таким образом, матрица $\mathbf{P_{\mu_1}}$ определяет марковскую цепь на дереве Кэли порядка $k^2$,
которое состоит из вершин исходного дерева в чётных местах.

Достаточное условие Кестена-Стигума не
крайности меры Гиббса $\mu_1$, соответствующей матрице
$\mathbf{P_{\mu_1}}$: $k^2s_2^2>1$, где $s_2$ есть
второе максимальное по абсолютной величине собственное значение $\mathbf{P_{\mu_1}}$.

Найдем собственные значения этой матрицы:
$$s_1=1, \ s_2=\frac{\lambda^2 z_1z_2}{\lambda^2 z_1z_2+\lambda z_1+\lambda z_2+1}.$$

\textbf{Случай $k=2$.} В этом случае
$$z_1=\left(\frac{\lambda+\sqrt{\lambda^2-4\lambda}}{2\lambda}\right)^2, \
z_2=\left(\frac{\lambda-\sqrt{\lambda^2-4\lambda}}{2\lambda}\right)^2.$$

В силу симметрии решений, достаточно проверить условие не крайности меры $\mu_1$ при $k=2$. Для этого вычислим
$\lambda^2 z_1z_2$ и $\lambda(z_1+z_2)$:
$$\lambda^2z_1z_2=\lambda^2\left(\frac{\lambda+\sqrt{\lambda^2-4\lambda}}{2\lambda}\right)^2\cdot\left(\frac{\lambda-\sqrt{\lambda^2-4\lambda}}{2\lambda}\right)^2$$

$$\lambda^2 z_1z_2=\lambda^2\left(\frac{\lambda^2-\lambda^2+4\lambda}{4\lambda^2}\right)^2=1$$

$$\lambda(z_1+z_2)=\lambda \frac{4\lambda^2-8\lambda}{4 \lambda^2}=\lambda-2$$

Тогда из $4s^2>1$ получим неравенство
$$4\left(\frac{1}{\lambda}\right)^2>1,$$
решение которого есть $\lambda<2$. Но меры $\mu_1$ и $\mu_2$ существуют при $\lambda>4$. Значит, эти меры заведомо являются  крайними.

Для исследования крайности приведем необходимые определения из работы \cite{MSW}.
Если удалить произвольное ребро $\langle x^0, x^1\rangle=l\in L$ из дерева Кэли $\Gamma^k$, то оно разбивается на две компоненты $\Gamma^k_{x^0}$ и $\Gamma^k_{x^1}$, каждая из которых называется полубесконечным деревом или полудеревом Кэли.

Рассмотрим конечное полное поддерево $\mathcal T$, которое содержит все начальные точки полудерева $\Gamma^k_{x^0}$. Граница $\partial \mathcal T$ поддерева $\mathcal T$ состоит из ближайших соседей его вершин, которые лежат в $\Gamma^k_{x^0}\setminus \mathcal T$. Мы отождествляем поддерево $\mathcal T$ с множеством его вершин. Через $E(A)$ обозначим множество всех ребер $A$ и $\partial A$.

В \cite{MSW} ключевыми являются две величины  $\kappa$ и $\gamma$. Оба являются свойствами множества мер Гиббса $\{\mu^\tau_{{\mathcal T}}\}$, где граничное условие $\tau$ фиксировано и $\mathcal T$ является произвольным, начальным, полным, конечным поддеревом $\Gamma^k_{x^0}$. Для данного начального поддерева $\mathcal T$ дерева $\Gamma^k_{x^0}$ и вершины $x\in\mathcal T$ мы будем писать $\mathcal T_x$ для (максимального) поддерева $\mathcal T$  с начальной точкой в $x$. Когда $x$ не является начальной точкой $\mathcal T$, через $\mu_{\mathcal T_x}^s$ обозначим меру Гиббса, в которой "предок"  $x$ имеет спин $s$ и конфигурация на нижней границе ${\mathcal T}_x$ (т.е. на $\partial {\mathcal T}_x\setminus \{\mbox{предок}\ \ x\}$) задается через $\tau$.

Для двух мер $\mu_1$ и $\mu_2$ на $\Omega$ через $\|\mu_1-\mu_2\|_x$ обозначим расстояние по норме
$$\|\mu_1-\mu_2\|_x={1\over 2}\sum_{i=0}^1|\mu_1(\sigma(x)=i)-\mu_2(\sigma(x)=i)|.$$
Пусть $\eta^{x,s}$ есть конфигурация $\eta$ со спином в $x$, равным $s$.

Следуя \cite{MSW}, определим
$$\kappa\equiv \kappa(\mu)={1\over2}\max_{i,j}\sum_{l=0}^1|P_{il}-P_{jl}|;$$
$$\gamma\equiv\gamma(\mu)=\sup_{A\subset \Gamma^k}\max\|\mu^{\eta^{y,s}}_A-\mu^{\eta^{y,s'}}_A\|_x,$$
где максимум берется по всем граничным условиям $\eta$, всеми $y\in \partial A$, всеми соседями $x\in A$ вершины $y$ и всеми спинами $s, s'\in \{0,1\}$.

Достаточным условием крайности  меры Гиббса $\mu$ является $k\kappa(\mu)\gamma(\mu)<1$,
но для рассматриваемых $G^{(2)}_k$-периодических мер это условие выглядит: $k^2\kappa(\mu)\gamma(\mu)<1$.

Используя (\ref{rus2.7}), при $i\neq j$ получим
$$\kappa=\frac{\lambda^2z_1z_2}{(1+\lambda z_1)(1+\lambda z_2)}.$$
А при $i=j$ имеем $|P_{il}-P_{jl}|=0$. Из работы \cite{MSW}(стр.151, Теорема 5.1.) известно, что для HC-модели справедлива оценка: $\gamma\leq{\lambda\over \lambda+1}$.

В случае $k=2$ для мер $\mu_1$ и $\mu_2$, соответствующих решениям $z_1$ и $z_2$, имеем $\kappa=\frac{1}{\lambda}$.
Следовательно, из условия $4\kappa\gamma>1$ получим неравенство
$$\frac{4\lambda}{\lambda(\lambda+1)}<1,$$
решением которого является $\lambda>3$. Следовательно, в случае $k=2$ условие крайности мер $\mu_1$ и $\mu_2$ выполняется при любых значениях $\lambda>4$, т.е. в области существования этих мер.

Итак, доказана следующая теорема.

\begin{theorem} Пусть $k=2$. Тогда для НС-модели $G^{(2)}_k$-периодические меры Гиббса
$\mu_1$ и $\mu_2$ при $\lambda>4$ являются крайними.
\end{theorem}

\textbf{Случай $k=3$.} В этом случае проверим условие не крайности меры $\mu^{(1)}$ (в силу симметрии решений и выражения для $s_2$ область не крайности меры $\mu^{(2)}$ совпадает с областью не крайности меры $\mu^{(1)}$). Из условия Кестена-Стигума $k^2s_2^2>1$ получим неравенство
$$h(\lambda)=9\left(\frac{\lambda^2z_1z_2}{(1+\lambda z_1)(1+\lambda z_2)}\right)^2-1>0.$$
Так как выражения для $z_1$ и $z_2$ громоздкие, решить это неравенство аналитически очень трудно. Поэтому рассмотрим производную
$h({1\over \lambda})$:
$$\left(h\left({1\over \lambda}\right)\right)'=-{1\over \lambda^2}h'\left({1\over \lambda}\right).$$
Ясно, что из возрастания функции $h({1\over \lambda})$ при $\lambda\in (0, 1]$ следует убывание функции $h(\lambda)$ при $\lambda \in [1, +\infty)$.
Из графиков функций $h(\lambda)$ и $h({1\over \lambda})$  можно увидеть, что функция $h(\lambda)$ убывает при $\lambda \in [1, +\infty)$ (см. Рис.2).  Кроме того, меры $\mu^{(1)}$ и $\mu^{(2)}$ существуют при $\lambda>\frac{27}{16}$.
Значит, эти меры заведомо являются крайними.

 \begin{center}
\includegraphics[width=6cm]{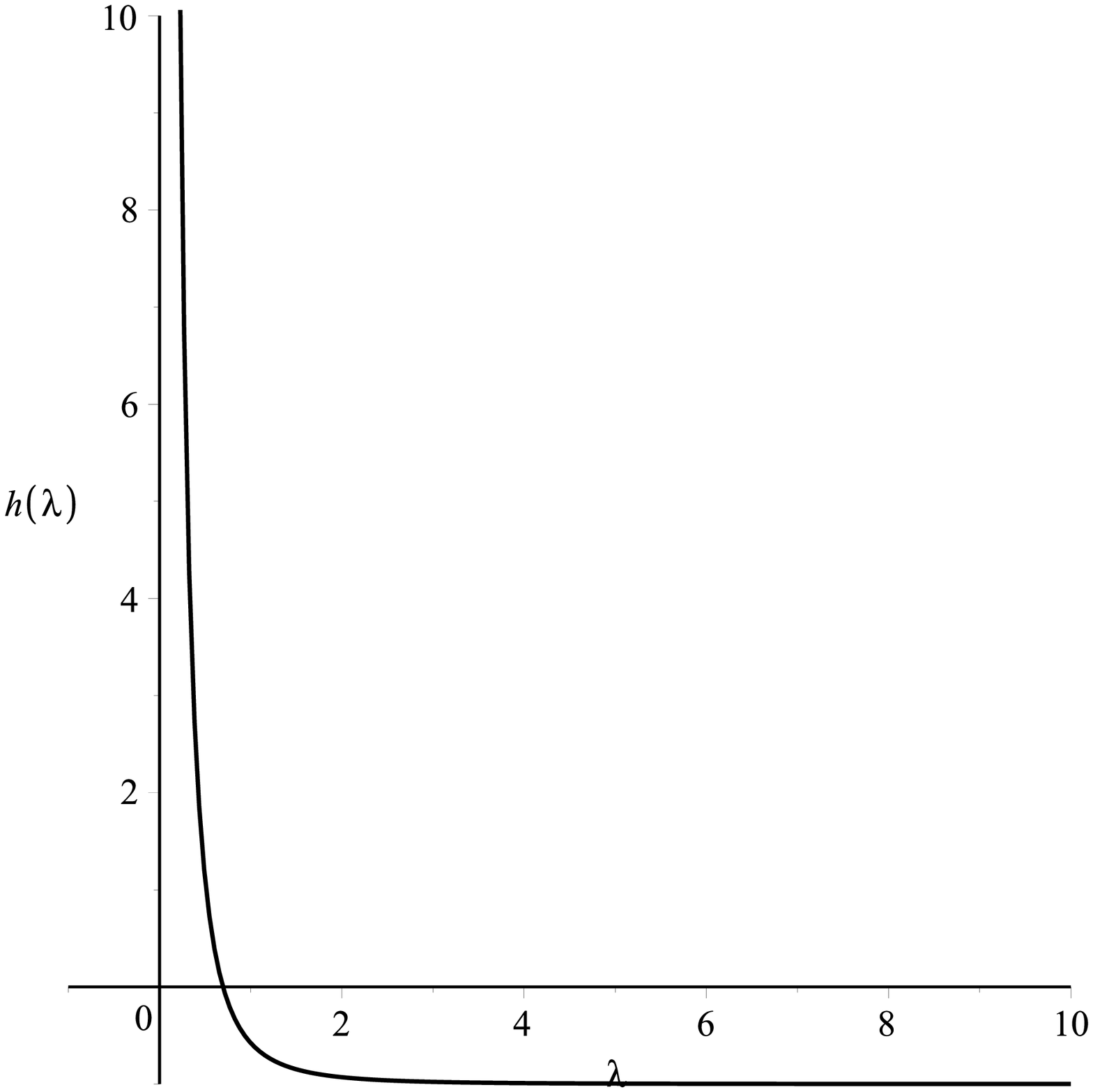} \ \ \  \includegraphics[width=6cm]{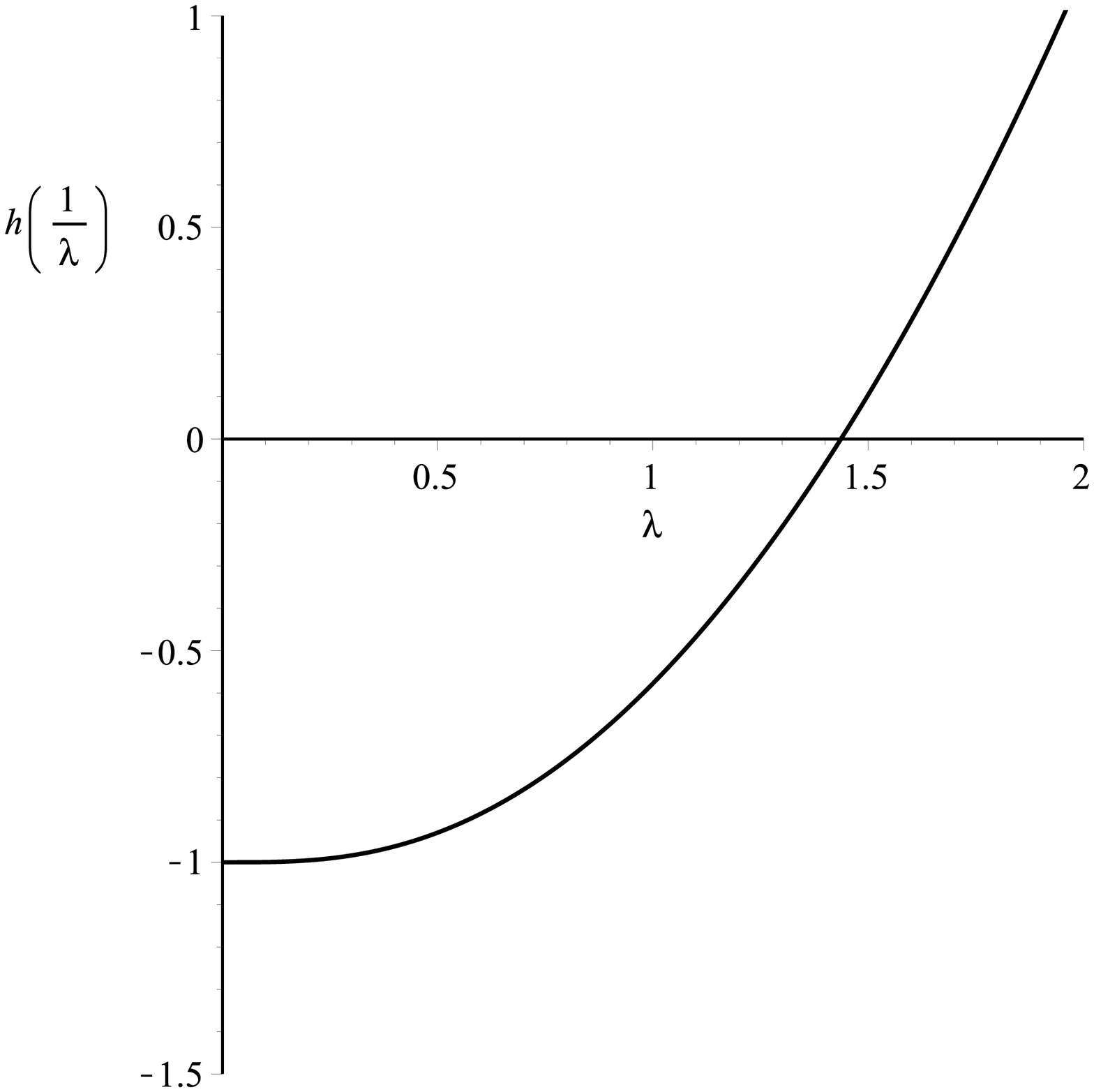}
\end{center}
\begin{center}{\footnotesize \noindent
 Рис. 2. График функции $h(\lambda))$ при $\lambda \in [1, +\infty)$ (слева) и график функции $h({1\over \lambda})$ при $\lambda\in (0, 1]$ (справа).}\
\end{center}

Далее, проверим условие крайности меры $\mu^{(1)}$ (в силу симметрии решений и выражения для $\kappa$ область крайности меры
$\mu^{(2)}$ совпадает с областью крайности меры $\mu^{(1)}$). Достаточным условием крайности меры $\mu^{(1)}$ является: $k^2\kappa(\mu^{(1)})\gamma(\mu^{(1)})<1$, т.е.
$$g(\lambda)=\frac{9\lambda^2z_1(\lambda)z_2(\lambda)}{(1+\lambda z_1(\lambda))(1+\lambda z_2(\lambda))}\cdot\frac{\lambda}{\lambda+1}-1<0.$$
Так как меры $\mu^{(1)}$ и $\mu^{(2)}$ существуют при $\lambda>\lambda_{cr}=\frac{27}{16}$, то неравенство $g(\lambda)<0$ можно рассмотреть при $\lambda \in [1, +\infty)$. Вычислим производную функции $g({1\over \lambda})$:
$$\left(g\left({1\over \lambda}\right)\right)'=-{1\over \lambda^2}g'\left({1\over \lambda}\right).$$
Из этого равенства и рисунка 3 следует, что функция $g(\lambda)$ убывает при $\lambda \in [1, +\infty)$, т.к. функция $g({1\over \lambda})$  возрастает при $\lambda\in (0, 1]$ (см. Рис.3). Следовательно, неравенство $g(\lambda)<0$ справедливо при $\lambda>\lambda_{cr}>1.$

\begin{center}
\includegraphics[width=6cm]{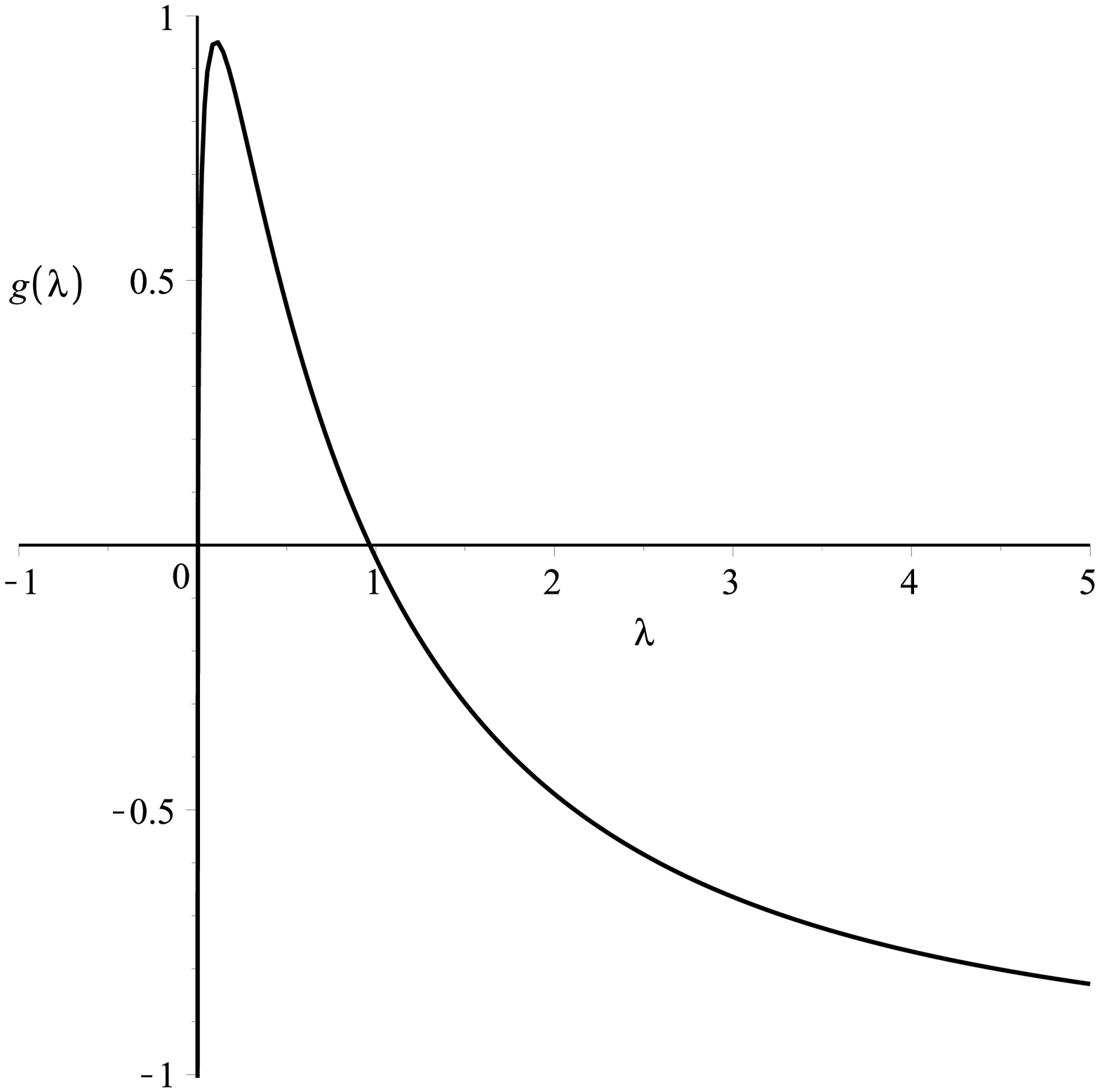} \ \ \ \  \includegraphics[width=6cm]{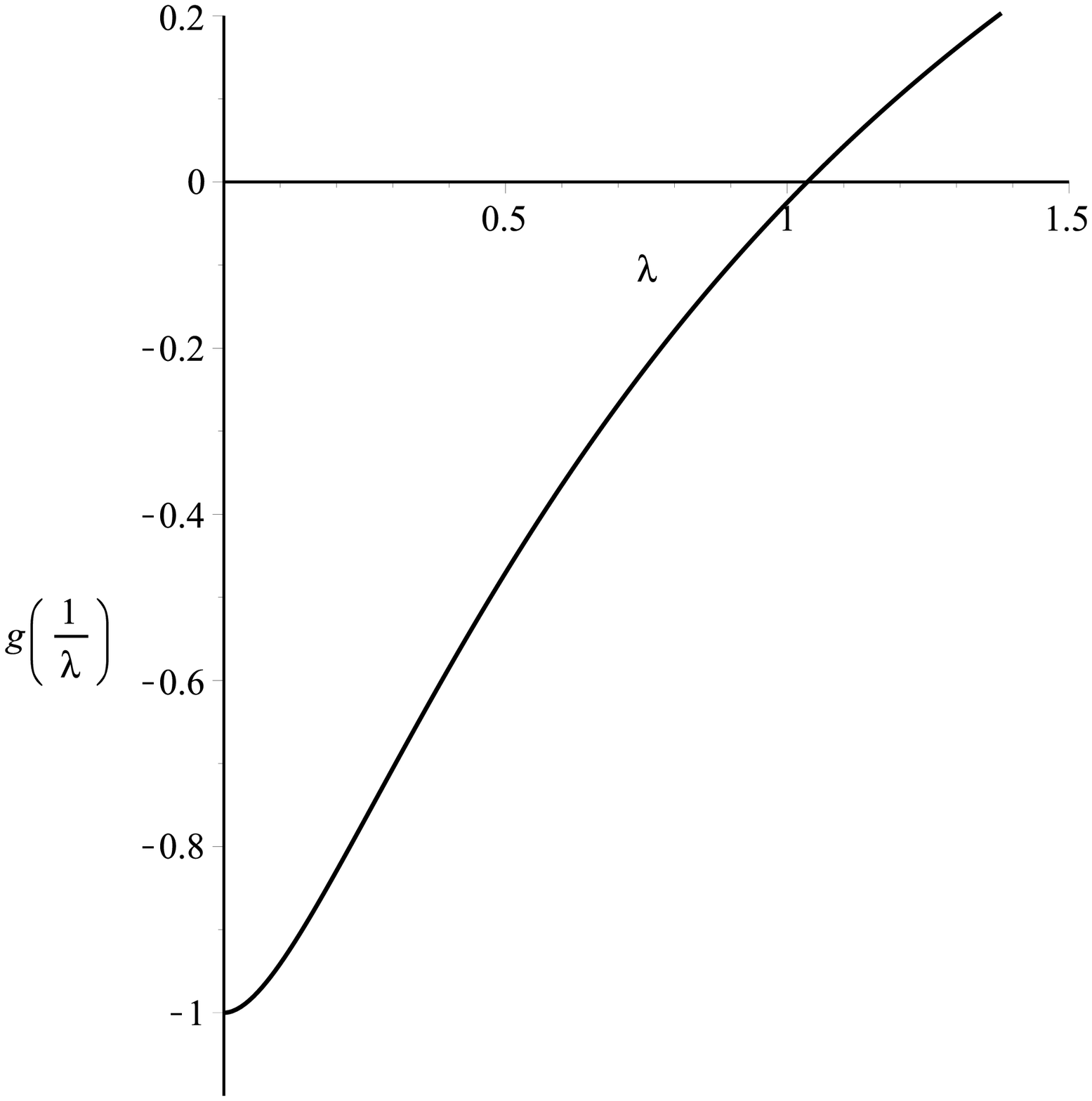}
\end{center}
\begin{center}{\footnotesize \noindent
 Рис. 3. График функции $g(\lambda)$ при $\lambda \in [1, +\infty)$ (слева) и график функции $g({1\over \lambda})$ при $\lambda\in (0, 1]$ (справа).}\
\end{center}

Итак, верна следующая теорема.

\begin{theorem} Пусть $k=3$. Тогда для НС-модели $G^{(2)}_k$-периодические меры Гиббса
$\mu^{(1)}$ и $\mu^{(2)}$ при $\lambda>\frac{27}{16}$ являются крайними.
\end{theorem}

\end{document}